\documentclass[pra,amsmath,twocolumn,showpacs]{revtex4-1}
\usepackage{graphicx}
\usepackage{amsmath}
\begin{document}
\def\la{{\langle}}
\def\ra{{\rangle}}
\def\vep{{\varepsilon}}
\newcommand{\beq}{\begin{equation}}
\newcommand{\eeq}{\end{equation}}
\newcommand{\beqa}{\begin{eqnarray}}
\newcommand{\eeqa}{\end{eqnarray}}
\newcommand{\q}{\quad}
\newcommand{\A}{\hat{A}}
\newcommand{\s}{\hat{S}}
\newcommand{\si}{\hat{\sigma}}
\newcommand{\Pp}{\hat{\Pi}}
\newcommand{\AC}{{\it AC }}
\newcommand{\La}{{\lambda }}
\newcommand{\n}{\\ \nonumber}
\newcommand{\om}{\omega}
\newcommand{\mi}{m_i^{\alpha}}
\newcommand{\mf}{m_f^{\alpha}}
\newcommand{\e}{\epsilon}
\newcommand{\Om}{\Omega}
\newcommand{\br}{\overline}
\newcommand{\cn}[1]{#1_{\hbox{\scriptsize{con}}}}
\newcommand{\sy}[1]{#1_{\hbox{\scriptsize{sys}}}}
\newcommand{\pd }{Pad\'{e} }
\newcommand{\PAD }{Pad\'{e}\q}
\newcommand{\PD }{Pad\'{e}\q}
\newcommand{\get }{\leftarrow}
\newcommand{\f}{\ref }
\newcommand{\R}{\text{Re } }
\newcommand{\I}{\text{Im } }

\title{Weak measurements measure probability amplitudes (and very little else)}
\author {D. Sokolovski$^{a,b}$}
\affiliation{$^a$ Departmento de Qu\'imica-F\'isica, Universidad del Pa\' is Vasco, UPV/EHU, Leioa, Spain}
\affiliation{$^b$ IKERBASQUE, Basque Foundation for Science, Maria Diaz de Haro 3, 48013, Bilbao, Spain}

\begin{abstract}
\noindent
Conventional quantum mechanics describes a pre- and post-selected system in terms of virtual (Feynman) paths via which the final state can be reached. In the absence of probabilities, a weak measurement (WM) determines the probability amplitudes 
for the paths involved. The weak values (VW) can be identified with these amplitudes, or their linear combinations.
This allows us to explain the "unusual" properties of the VW, and avoid the "paradoxes" often associated with the WM.  

\date{\today}
\end{abstract}
\maketitle

\vskip0.5cm
\date{\today}
\section{introduction}
Evers since inception in 1988 \cite{Ah} the subject of the so-called quantum  weak values  (WV) remained a controversial topic
(for early critique see \cite{C1}, \cite{C2}). In more recent developments, the authors of \cite{CRAP} put forward a controversial  \cite{COMM}, \cite{DS1} of generalising the WV to classical theories,
 while Steinberg  \cite{Stein} suggested that  
weak measurements (WM) can be used for probing certain "surreal" elements of quantum physics.  Recently, the authors of \cite{Nature} have demonstrated experimentally how WM can be used to (indirectly) measure the system's wave function. One might feel that  a clarification of what actually happens in a WM is in order, and the purpose of this paper is to provide one based on the concepts conventionally used in quantum theory.

The history of WV goes back to Feynman, who used mean value of a functional,  averaged with the {\it probability amplitudes}, to illustrate certain aspects of quantum motion \cite{FEYN1}. Feynman averages naturally arise, for example, in an attempt to measure 
time spent by a tunnelling particle in the barrier \cite{SB}.
The WM, designed to perturb the measured system is little as possible, were later studied in terms of Krauss operators and the POVM's, and found applications in the analysis of continuous measurements \cite{Mensky1}, \cite{Caves}, \cite{Mensky2}. 
The subject gained in popularity when the authors of \cite{Ah} pointed out some "unusual" properties of the VW.
Subsequent attempts to better understand  these properties were made, for example, in the analysis of the "complex probabilities"
in \cite{Hoff}.
In a recent review of the practical aspects of VW \cite{Rev} the authors characterised the VW as {\it "complex numbers that one can assign 
to the powers of a quantum observable operator  $\hat{A}$ using two states, and initial state $|i\ra$...., and a final state $|f\ra$..."}
This still leaves open the original question posed by the authors of \cite{Ah}: what, if anything, the WV tell us about the intermediate state of a pre- and post-selected system? 

We will answer it in the following way: in  the case of intermediate measurements  on a pre- and post-selected system one must consider the system's histories referring to at least three different  moments of time. Such histories, in general, interfere, and are conventionally  characterised by probability amplitudes \cite{FEYN}. A WM destroys coherence between the histories only slightly and, in the absence of probabilities, measures the corresponding probability amplitudes or, more generally, various combinations of its real and imaginary parts.
We will show that this simple observation allows one to avoid the notions of  "anomalous" weak values \cite{Ah}, \cite{CRAP}, quantum system "being at two different places at the same time" \cite{Ahbook}, "photons disembodied from its polarisation" \cite{CAT},\cite{CAT2}, or violation of Einstein's causality in classically forbidden transitions \cite{NIM}.
For consistency, we will need to reproduce some of the known results, and we will try to do it in the briefest possible manner in the following Sections.
\section{Paths, amplitudes, and meters}
Following \cite{Ah} we consider a system in an $N$-dimensional Hilbert space with a Hamiltonian $\hat{H}$. We also 
consider an arbitrary operator $\s$, with the eigenvalues $S_i$ and the eigenstates $|i\ra$, $i=1,2,,..,N$.
At $t=0$ the system is prepared (pre-selected) in a state $|\psi\ra$
and at $t=T$ we check if the system is (post-select the system) in another state $|\phi\ra$.
If it is, we will keep the results of all other measurements we may make  halfway into the transition, at $t=T/2$.
The probability amplitude for a successful post-selecton is then $A^{\phi \gets \psi}=\la \phi|\exp(-i\hat{H}T)|\psi\ra$.
Inserting the unity $\sum_{i}|i\ra\la i|=1$ at $t=T/2$ we have
 \begin{eqnarray}\label{a3}
A^{\phi \gets \psi}=\sum_{i=1}^NA^{\phi \gets \psi}_i, \q\q\q\q\q\q\n
 A^{\phi \gets \psi}_i\equiv \la \phi|\exp(-i\hat{H}T/2)|i\ra\la i|exp(-i\hat{H}T/2)|\psi\ra
\end{eqnarray}
This can be seen as a variant of the most basic quantum mechanical problem  \cite{FEYN}: a system may reach the final state 
from the initial state 
 via $N$ paths or path (see Fig. 1). The paths are determined by the nature of the quantity $\s$, and their amplitudes depend on $\s$, as well as on the initial and final states $|\psi\ra$ and $|\phi\ra$.
\newline
The paths may be either interfering  or exclusive alternatives \cite{FEYN}, depending on what is done at $t=T/2$.
If nothing is done, $N$ {\it virtual} paths form a single route, and 
their amplitudes should be added as in Eq.(\ref{a3}) \cite{FEYN}. The probability to arrive in $\phi$ is then given by $P^{\phi \gets \psi}=|\sum_{i=1}^NA^{\phi \gets \psi}_i|^2$. Alternatively, an external meter can destroy interference between the paths.
If the destruction is complete, the paths become {\it real} and can be equipped with probabilities $|A^{\phi \gets \psi}_i|^2$.
The probability of a successful post-selection is now given by $P^{\phi \gets \psi}=\sum_{i=1}^N|A^{\phi \gets \psi}_i|^2$.
\begin{figure}
	\centering
		\includegraphics[width=8cm,height=5cm]{{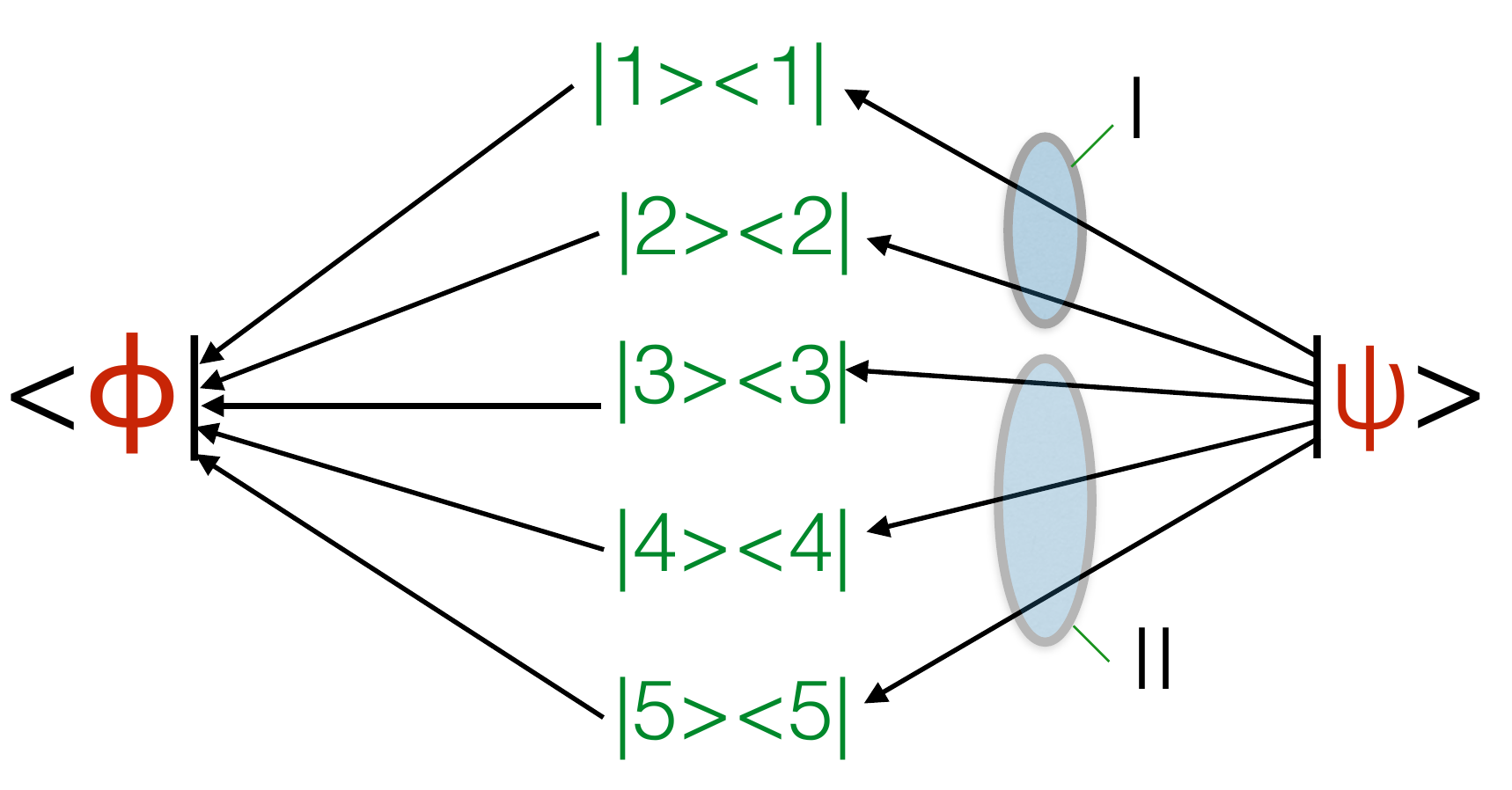}}
\caption{(Color online) 
A system in a $5$ dimensional Hilbert space can reach the final state $|\phi\ra$ via five virtual paths $\{i\}$ 
with probability amplitudes $ A^{\phi \gets \psi}_i$.
An accurate measurement of an operator $\s=\sum_{i=1}^2|i\ra\la i|-\sum_{i=3}^5|i\ra\la i|$ with degenerate eigenvalues of $1$ and $-1$
creates two real pathways $I=\{1+2\}$ and $II=\{3+4+5\}$, travelled with the probabilities $\omega_{I}$
and $\omega_{II}$ given by Eq.(\ref{a7aaa}).
A WM of $\s$ determines the difference relative amplitudes for the virtual paths $I$ and $II$ in Eq.(\ref{a9a}), $\alpha_{I}-\alpha_{II}$. 
}
\label{fig:3}
\end{figure}
\section{Von Neumann measurements with post-selection}
To see how interference between the paths  shown in Fig. 1 can be destroyed, we employ a von Neumann pointer with a position $f$ and momentum  $\lambda$, briefly coupled to the system around $t=T/2$ via an interaction Hamiltonian $-i\delta(t-T/2)\partial_f \s $ (we use $\hbar=1$). The meter is prepared in a state $|M\ra$, 
such that $G(f)\equiv \la f|M\ra$ is a real function which peaks around the origin $f=0$ with a width $\Delta f$, 
\begin{eqnarray}\label{a4c}
G(f)=\la f| M\ra=(\Delta f)^{-1/2}G_0(f/\Delta f)
\end{eqnarray}
where $G_0(f)=G_0(-f)$, $G_0(f)_{|f|\to \infty}\to 0$ and $\int G_0^2(f)df=1$. After a successful post-selection, the meter is in a pure state $|M'\ra$  (the result is well known, see, for example, \cite{Ah})
\begin{eqnarray}\label{a4a}
G'(f)=\la f| M'\ra= \sum_{i=1}^NA^{\phi\gets\psi}_iG(f-S_i).
\end{eqnarray}
In the momentum space, the meter's final state is given by
\begin{eqnarray}\label{a4b}
G'(\lambda)=\la \La| M'\ra= G(\lambda) \sum_{i=1}^NA^{\phi\gets\psi}_i\exp(-i\lambda S_i),
\end{eqnarray}
where $G'(f)=(2\pi)^{-1/2}\int G(\lambda)\exp(i\lambda f)d\lambda$.
Repeating the experiment many times we can evaluate the mean pointer position or the  momentum after the measurement, 
\begin{eqnarray}\label{a4}
\la f\ra_{\s}= \int f |G(f)|^2 df/ \int |G(f)|^2 df,
\end{eqnarray}
and 
\begin{eqnarray}\label{a4e}
\la \lambda \ra_{\s}= \int \lambda |G(\La)|^2 d\La/ \int |G(\La)|^2 d\La.
\end{eqnarray}
So what can be learnt about the condition of a pre- and post-selected system at $t=T/2$?
It is convenient to write the operator $\s$ as a sum of projectors on its eigenstates,
 \begin{eqnarray}\label{a5}
\s=\sum_{i=1}^N S_i\hat{P}_i, \q \hat{P}_i\equiv |i\ra\la i|.
\end{eqnarray}
and consider  the measurement of a $\hat{P}_i$ for various values of $\Delta f$.
\section{Accurate (strong) measurements}
Consider first an accurate (strong) measurement of a $\hat{P}_i$. Since $\Delta f$ determines the uncertainty in the initial setting of the pointer, an accurate measurement would require $\Delta f\to 0$. If so, we easily find that
\begin{eqnarray}\label{a6}
\la f\ra_i^{strong}= |A^{\phi\gets\psi}_i|^2/|\sum_{i'=1}^N|A^{\phi\gets\psi}_{i'}|^2\equiv \omega_i.
\end{eqnarray}
Thus, an accurate meter completely destroys the coherence between the paths in Fig.1. Moreover, the measured mean value of the projector $\hat{P}_i$ gives the {\it relative frequency} with which the real path passing through the  $i$-th state is travelled if the experiment is repeated many times. It is a simple matter to verify that for an arbitrary operator $\s$  with non-degenerate eigenvalues, $S_i\ne S_j$, 
the mean value of the pointer position gives the weighted sum of its  eigenvalues, 
\begin{eqnarray}\label{a7}
\la f\ra^{strong}_{\s}= \sum_{i=1}^N\omega_iS_i.
\end{eqnarray}
This has an obvious classical meaning: if the value of the quantity $\s$ on the $i$-th path is $S_i$, and the $i$-th path is travelled 
with the probability $\omega_i$, then the average value over many trials is given by by the sum (\ref{a7}).
\newline
If $K$ and $(N-K)$ eigenvalues of the measured $\s$ are degenerate, e.g., $S_{1}=...=S_{K}\equiv S_I$, $S_{K+1}=...=S_{N}\equiv S_{II}$, Eq.(\ref{a4a})  shows that the interference between the paths within each group of eigenvalues are not destroyed by a strong measurement (SM) of $\s$ (see Fig. 1). Rather, in accordance with the Uncertainty Principle \cite{FEYN},\cite{DS2}, 
they are combined into two real routes, with amplitudes,
\begin{eqnarray}\label{a7aa}
A^{\phi\gets\psi}_{I} =\sum_{i=1}^KA^{\phi\gets\psi}_{i},\q \text{and} \q A^{\phi\gets\psi}_{II} =\sum_{i=K+1}^NA^{\phi\gets\psi}_{i},\q
\end{eqnarray}
which are  travelled with the probabilities 
\begin{eqnarray}\label{a7aaa}
\omega_{I}=|A^{\phi\gets\psi}_{I}|^2/(|A^{\phi\gets\psi}_{I}|^2+|A^{\phi\gets\psi}_{II}|^2), \n
\omega_{II}=|A^{\phi\gets\psi}_{I|}|^2/(|A^{\phi\gets\psi}_{I}|^2+|A^{\phi\gets\psi}_{II}|^2).
\end{eqnarray}
Accordingly, we have
\begin{eqnarray}\label{a7a}
\la f\ra^{strong}_{\s}= \omega_{I} S_I+ \omega_{II} S_{II}.
\end{eqnarray}
Equation (\ref{a7a}) is easily generalised to the case where there are more than two groups of degenerate eigenvalues.
\newline
Thus, a strong measurement measures probabilities (and their linear combinations) for the real paths created by the meter according to the simple rules (\ref{a7}) and (\ref{a7a}). Importantly, different choices of the operator $\s$ lead to different sets of the real paths and, therefore, to different statistical ensembles which may have nothing more in common than their parent quantum system \cite{DS2}. We will return to this point in Sections IX and X.
\section{Inaccurate (weak) measurements}
Suppose next that we want to know something about the quantity $\s$ at $t=T/2$, provided all paths are allowed to interfere. The Uncertainty Principle may warn us against such an attempt \cite{DS1}, \cite{DS2}, but we want to proceed anyway, 
by making the measurement highly inaccurate, or weak. We send $\Delta f \to \infty$, so that the r.h.s of  Eq.(\ref{a4a}) becomes 
$G(f)= const\times \sum_{i=1}^NA^{\phi\gets\psi} +\delta$, where $\delta$ is a small correction which would, hopefully, tell us something about the system when we measure $\la f \ra_{\s}$.  But what precisely? With the interference almost untouched, we have no probabilities for the virtual paths, yet there are always probability amplitudes. We would risk a guess: $\la f \ra_{i}$ will tell us something about the { probability amplitude} to travel the $i$-th path. 
\newline
As before, we will proceed by trying to measure the projectors  $\hat{P}_i$ in the limit of vanishing accuracy, $\Delta f \to \infty$, and determine the mean pointer reading $\la f\ra_{i}$. Evaluating, to the first order of  $\partial_fG$, the average in Eq.(\ref{a4})
we have 
\begin{eqnarray}\label{a8}
\la f\ra^{weak}_i= \R \alpha_i,
\end{eqnarray}
where 
\begin{eqnarray}\label{a8a}
\alpha_i\equiv 
 \frac{A^{\phi\gets\psi}_i}{\sum_{i'=1}^NA^{\phi\gets\psi}_{i'}}.
 \end{eqnarray}
We note that Eq.(\ref{a8}) has the same structure as Eq.(\ref{a6}), but with probabilities replaced with the real parts of the corresponding  {\it relative probability amplitudes} $\alpha_i$. Similarly, the mean reading of a weak meter set to measure an arbitrary operator $\s$ is just 
the weighted sum of its eigenvalues,  
\begin{eqnarray}\label{a9}
\la f\ra^{weak}_{\s}= \sum_{i=1}^NS_i \R \alpha_i.
\end{eqnarray}
For $\hat{H}=0$, this can be re-written in an equivalent form, used, for example, in \cite{Ah}, 
\begin{eqnarray}\label{a9}
\la f\ra^{weak}_{\s}= \sum \la \phi|i\ra \hat{A}\la i|\psi\ra/\la\phi|\psi\ra=\frac{\la \phi|\hat{A}|\psi\ra}{\la\phi|\psi\ra}.
\end{eqnarray}
where 
Since there are no {\it a priori} restrictions on the sign of $\R \alpha_i$, Eq.(\ref{a9}) has no simple probabilistic interpretation,
\cite{DS3} and $\la f\ra^{weak}_{\s}$ may take any real value at all. We note that Eq.(\ref{a9}) is valid whether or not the eigenvalues of $\s$ are degenerate. For example, for an operator with two sets of degenerate eigenvalues, $S_{I}$ and $S_{II}$, discussed in Sect. IV, Eq.(\ref{a9}) gives
\begin{eqnarray}\label{a9a}
\la f\ra^{weak}_{\s}= S_I\text{Re } \alpha_I+S_{II}\text{Re }\alpha_{II},\q \n 
\alpha_{I,II}=A^{\phi \gets \psi}_{I,II}/(A^{\phi \gets \psi}_{I}+A^{\phi \gets \psi}_{II}).
\end{eqnarray}
\newline
To measure the imaginary parts of the $\alpha$'s,
we follow \cite{Ah}, and look at the mean momentum acquired by the pointer given by Eq.(\ref{a4e}). 
For the projector $\hat{P}_i$, whose only non-zero eigenvalue is $S_i=1$, Eq.(\ref{a4b}) reduces to
\begin{eqnarray}\label{a10}
G'(\lambda)= G(\lambda) [\sum_{j\ne i}A^{\phi\gets\psi}_{j}+A^{\phi\gets\psi}_i\exp(-i\lambda )].
\end{eqnarray}
With $G(f)$ broad in the co-ordinate space, $G(\lambda)$ is narrow. Thus, $\exp(-i\lambda )\approx 1-i \lambda$, and using
Eq.(\ref{a4b}), for the projector $\hat{P}_i$  and, for an arbitrary quantity $\s$ we have
\begin{eqnarray}\label{a11}
\la \lambda \ra^{weak}_{i}= 2\int \lambda^2 |G(\lambda)|^2 d\La \times \I \alpha_i, 
\end{eqnarray}
and 
\begin{eqnarray}\label{a12}
\la \lambda \ra^{weak}_{\s}= 2\int \lambda^2 |G(\lambda)|^2 d\La
 \times \sum_{i=1}^N  S_i\I \alpha_i, 
\end{eqnarray}
respectively.
Therefore, a weak von Neumann meter can be used to completely determine the complex valued amplitudes $\alpha_i= \R \alpha_i+i\I \alpha_i$.  If we cannot measure the projectors $\hat{P}_i$ directly, but can do so for a set of N operators $\s^j$ such that the matrix $S^j_i$ is non-degenerate,  the $2N$ measured mean readings,  
$\la f \ra^{weak}_{\s^j}$ and $\la \La \ra^{weak}_{\s^j}$ can be used to reconstruct all amplitudes $\alpha_i$.
\newline
Thus, an inaccurate meter does nor create new real pathways, and measures instead the
 {probability amplitudes}  for the virtual paths shown in Fig.1. 
Confusing probabilities with amplitudes  may lead to "paradoxes", but it would be wrong to attribute them the quantum theory, as will be discussed below. 
\section {Tomography of a transition}
Since a weak meter  perturbs a transition only slightly, several WM can be performed sequentially, or even at the same time \cite{AhM}. To check this we can use the close relation between the accuracy and the perturbation incurred in a measurement.
Let us measure operators $\s^{(j)}$, $j=1,2,..,J$, which may or may not commute, at the same time $t=T/2$, and with all pointers prepared in the same state (\ref{a4c}). The coupling Hamiltonian is now  $\hat{H}_{int}=-i\delta(t-T/2)\sum_{j=1}^J \partial_{f_j},\s^{(j)}$ and changing the variables $f'_j=f_j/\Delta f$, we may rewrite it as $-i(\Delta{f})^{-1}\delta(t-T/2)\sum_{j=1}^J \partial_{f'_j}\s^{(j)}$. 
Determination of all meter positions and momenta gives the results $\overline{f}\equiv(f_1,f_2,...,f_N)$, and $\overline{\La}\equiv(\La_1,\La_2,...,\La_N)$, respectively.

Making all measurements weak by sending $\Delta f \to \infty$, for  the evolution  operator over the time of interaction with the meters we obtain
\begin{eqnarray}\label{d1}
\exp(-i \hat{H}_{int}/\Delta f)\approx 1-\frac{1}{\Delta f}\sum_{j=1}^J \partial_{f'_j}\s^{(j)}.
\end{eqnarray}
Using (\ref{d1}) to evolve the initial state $(\Delta f)^{K/2}\prod_j G_0(f'_j)|\psi\ra $, projecting the result on the final state $|\phi\ra$, and returning to the original variables $f_j$, we find the (yet unnormalised) final state of the pointers,
\begin{eqnarray}\label{d2}
G'(\overline{f})=\prod_j\la f_j| M'\ra= \q\q\q\q\q\q\q\q\q\n
\prod_{j=1}^K G(f_j)\left [\la \phi|\psi\ra- \sum_{j=1}^K\frac{\partial_{f_j}G(f_j)}{G(f_j)}]\la \phi|\s^{(j)}|\psi\ra\right ].
\end{eqnarray}
Evaluating, to the first order of $\partial_{f_j}G(f_j)$, the mean pointer positions gives then the same result as a WM
of the operator $\s^{(j)}$ alone in Eq.(\ref{a9}) ($d\overline{f}=\prod_i df_i$), 
\begin{eqnarray}\label{d3}
\nonumber
\int f_j \rho_f(\overline{f})d\overline{f}/\int  \rho_f(\overline{f})d\overline{f}=
\sum_{i=1}^NS^{(j)}_i \R \alpha^{(j)}_i,
\end{eqnarray}
where $\rho(\overline{f})\equiv |G'(\overline{f})|^2$, and  $\alpha^{(j)}_i$ is the relative amplitude for the $i$-path, defined in the representation in which operator $\s^{(j)}$ is diagonal with the eigenvalues $S^{(j)}_i$.
Similarly, for the mean momentum of the $j$-th meter we recover the result (\ref{a12}),
\begin{eqnarray}\label{d4}
\nonumber
\int \La_j \rho_\La (\overline{\La})d\overline{\La}/\int\rho_\La (\overline{\La})d\overline{\La}
\q\q\q\q\q\q
\\
= 2\int \lambda^2 |G(\lambda)|^2 d\La
 \times \sum_{i=1}^N S^{(j)}_i\I \alpha^{(j)}_i,\q\q
\end{eqnarray}
where $\rho_\La(\overline{\La})\equiv |G'(\overline{\La})|^2$, and $G'(\overline{f})$ $=(2\pi)^{-K/2}\int d\overline{\La}_K$ $ G'((\overline{\La}) \exp(i\sum_j \La_jf_j)$.
\newline
One way to use simultaneous WM is to employ, at the same time, $N$ pairs of meters, each pair measuring one of the orthogonal projectors in the  same representation.  
 $\hat{P}_{j}=|j\ra \la j|$. Determining all mean positions and mean momenta, we will then have a complete set of the path amplitudes $\alpha_i$ in Eq.(\ref{a8a}). We can repeat the experiments using different basis, thus performing a complete "tomography" of the transition between the states $|\psi\ra$ and  $|\phi\ra$. One can also simultaneously measure the amplitudes for different basis' 
 (provided, of course, that the joint effect of all the meters involved on the studied transition is small).
\newline
 The knowledge of all relative amplitudes $\alpha_i$ allows one to predict the results of any strong measurement without actually making it. In the simplest case where none of the eigenvalues of $\s$ are degenerate, the ratios of the frequencies $\omega$ in Eq.(\ref{a7}) are readily expressed in terms of the $\alpha$'s,  $\nu_i\equiv\omega_i/\omega_N=|\alpha_i|^2/|\alpha_N|^2$, $i=1,2,...,N-1$. The first $N-1$ probabilities $\omega_i$ are then found by solving $N-1$ linear equations, 
\begin{eqnarray}\label{d5}
\sum_{i=1}^N (1+\delta_{ij}/\nu_j)\omega_i=1, \q j=1,2,...,N-1,
\end{eqnarray}
 and the remaining $\omega_N$ is just $1-\sum_{i=1}^{N-1}\omega_i$.
 The case of degenerate eigenvalues can be  treated in a similar manner, and we will not go into details here.
 \newline
There is no conflict with the Uncertainty Principle: virtual amplitudes for all possible paths are potentially present in the unperturbed transition, just as the projections on all possible basis are potentially present in he  state $|\psi\ra$ describing the system at a given time.
Nothing of the above is unusual. From the first-order perturbation theory it is well known that a response of a quantum system to a small perturbation contains information about both the moduli and phases of the amplitudes involved (see, for instance, \cite{DSC}).
\section{Interpretation of the weak values}
As was discussed in Sect. IV, a strong mean value of the projector $\hat{P}_i$ gives us the frequency with which the $i$-th real route, created by the meter, 
would be travelled if the experiment is repeated many times. A weak mean value of  $\hat{P}_i$ in Eq.(\ref{a11}), on the other hand, simply tells us what the real or imaginary part of the amplitude $\alpha_i$ is. The same applies to any operator $\s$. The quantity $\la f\ra^{weak}_{\s}$ is just a weighted sum of the real parts of $\alpha_i$. It is a simple matter to show that for any  $|\psi\ra$, and $N$ complex quantities  $z_1$, $z_2$, ...$z_N$ adding to unity, $\sum_i z_i=1$, one can always find $|\phi\ra$ such that $\alpha_i= z_i$. 
Indeed, equating $\alpha_i$ to $z_1$ gives us a set of linear equations.
\begin{eqnarray}\label{a10a}
\sum_{i=1}^N(z_j-\delta_{ij})A^{\phi\gets\psi}_i=0, \q j=1,..,N
\end{eqnarray}
which can always be solved for $A^{\phi\gets\psi}_i=\la\phi |i \ra \la i|\psi \ra $ and, therefore for $\phi$, provided all $z_i$ add up to unity.
By the same token, one can always find a pre- and post-selected system such that the measured weak value (\ref{a9})
will be  {\it any} real number, large or small, positive or negative. In Ref.\cite{DS1} we linked this to the Uncertainty Principle, which forbids dividing interfering alternatives into sets which have individual physical significance.
Now "any" means that there would be values well outside the spectrum of the operator $\s$, and this is not surprising.
There will also be values inside the spectrum, but this is not a rule.
Finally, there must be a value exactly the same as the one we would obtain in the strong measurement, 
$\la f\ra^{weak}_{\s}=\la f\ra^{strong}_{\s}$,
but it has no special significance. (It is easy to check that this would be the case for a system post-selected in 
$|\phi\ra = \exp(-i\hat{H}T)|\psi\ra$).
\newline
It is our central argument that one shouldn't look for an interpretation more meaningful than the one given above.  
For example, it would be unwise to interpret $\la f\ra^{weak}_{\s}$ as the mean value of $\s$ under the conditions where the interference is not destroyed.
Such an interpretation would lead to an intriguing but wrong conclusion that quantally, and for reasons unknown,  any physical quantity can take a huge value, and complicate the issue, rather than simplify it.  
Thus, anyone measuring weakly, say, the mass of an electron, and finding a value of $10^{10}$ kg, (let alone $-10^{10}$ kg) would need to account for the absence of a massive gravitational field such a mass would produce. 
 Not even the innocent looking $\la f\ra^{weak}_{\s}=\la f\ra^{strong}_{\s}$ can be interpreted in the same way as $\la f\ra^{strong}_{\s}$ itself. Like all other weak values, it is obtained under different physical conditions, where the meter measures amplitudes rather than probabilities. 
 Recognising  $\la f\ra^{weak}_{\s}$ as nothing more than a weighted combination of probability amplitudes, which can take any value by their very nature, returns the discussion to the realm of the  reasonable.
 Next we will illustrate this by looking at some of the most often discussed weak measurement "paradoxes", simplifying the narrative where possible.
 
 \section{How can a measurement of a spin $1/2$ give a result $100$?}
 Following \cite{Ah}  we consider a weak measurement of the $z$-component of a spin-$1/2$, $\s=\sigma_z=|1\ra\la1|-|2\ra\la2|$.
 The spin is pre- and post-selected in the states
 \begin{eqnarray}\label{b1}
\psi=(|1\ra+|2\ra)/\sqrt{2}, \q \phi=(|1\ra+b|2\ra)/\sqrt{1+|b|^2},
\end{eqnarray}
and we put $\hat{H}=0$, so that nothing happens before and after the WM is made. There are, therefore, two virtual paths, with the relative probability amplitudes of $\alpha_1=1/(1+b^*)$ and $\alpha_1=b^*/(1+b^*)$, respectively.
\newline
Let us choose $b$ to be real, $b=-99/101$. It is easy to see that the mean reading of the weak meter will equal $100$, 
 \begin{eqnarray}\label{b2}
\la f\ra^{weak}_{\sigma_z}=100.
\end{eqnarray}
What can we infer from  this simple and verifiable result?
The authors of  \cite{Ah}  conclude that  {\it "... the usual measuring procedure for preselected and post-selected ensembles of quantum systems gives unusual results. Under some natural conditions of weakness of the measurement, its result consistently defines a new kind of value for a quantum variable, which we call the weak value."}
\newline
Recalling that $\s =\sigma_z$ has eigenvalues $s_1=1$ and $s_2=-1$, and consulting with Eq.(\ref{a9}), we see, however, that the  "new kind of quantum variable",  in this case, is just the difference between the real parts of the corresponding amplitudes, which indeed equals $100$,  $Re \alpha_1-Re \alpha_2 = Re(1-b^*)/(1+b^*)=100$. There is nothing "unusual" about this result.
 \section{The "quantum Cheshire cat"} 
 The authors of \cite{CAT} consider a system consisting of two parts, each described by a variable taking two possible values.
 The particle (the cat) can be found in the state $|L\ra$ (on the left), or in the state $|R\ra$ (on the right). The particle carries a spin
 (or the cat carries a smile) whose projections  on the $z$-axis may take values $\pm 1$.  Strong measurement (SM) of the operator $\Pp_R=|R\ra \la R|$ establishes whether the cat is on the right, without asking questions about the spin.
 Strong measurement of the projectors $\si^+_R= \Pp_R|+\ra \la +|$ and  $\si^-_R= \Pi_R|-\ra \la -|$ checks whether the cat is on the right, with its spin up and down, respectively. The operator $\si_{L,R}=(\si^+_{L,R}-\si^-_{L,R})$ indicates, according to \cite{CAT}, the presence of angular momentum at the corresponding location.
 The authors of \cite{CAT} show that there exist initial and final states such that an intermediate SM of $\Pp_R$ at $t=T/2$ give (it is also assumed that $\hat{H}=0$)
  \begin{eqnarray}\label{b2a}
\la f\ra^{strong}_{\Pp_R}=0,
\end{eqnarray}
while a SM of $\si_{R}$ registers the presence on the right  of the particles with the spin pointing up and down, with the probabilities 
$\omega_3$ and $\omega_4$, respectively.
Concluding from (\ref{b2a}) that the particle (the photon in the optical realisation of the experiment) must always pass through the left arm of the apparatus, yet seeing the angular momentum (polarisation) in the right arm, the authors of \cite{CAT} suggest that they {\it "... know with certainty that the photon went through the left arm, yet  find angular momentum in the right arm". } 
They continue to assert that 
 {\it "... physical properties can be disembodied from the objects they belong to."}. 
 \newline
 We might argue against these  conclusions by pointing out that in a SM of $\Pi_R$ a particle, which is indeed on the left, still carries its spin in a state $a_+|+\ra+a_-|-\ra$, $a_\pm=\la R|\la \pm|\psi\ra$, polarised along a direction other then the $z$-axis.  Thus, the smile has not completely left the cat, but just moved to some other part, perhaps to the cat's back. Similarly, $\omega_3+\omega_4$ gives the total probability for the particle with a known $z$-projection of its spin to pass through the right arm, so the smile there is not entirely without a cat.
\newline
It can also be argued differently. In the chosen representation, there are four virtual paths connecting the two states: $\{1\}$ with the cat on the left and the spin up, $\{2\}$ with  the cat on the left and spin down, 
$\{3\}$ with the cat on the right and the spin up, and $\{4\}$, with the cat on the right and the spin down.  There are also  four 
relative amplitudes $\alpha_i$, $i=1,2,3,4$.
Consulting with Sect. IV we note that a SM of $\Pp_R$ creates two real paths, $\{1+2\}$ comprising the paths $\{1\}$ and $\{2\}$, and
$\{3+4\}$, consisting of the paths $\{3\}$ and $\{4\}$. Since the second real path is not travelled, we know that  $\alpha_4=-\alpha_3$.
A SM of the operator  $\si_R$ creates three real paths,  $\{1+2\}$ corresponding to its doubly degenerate eigenvalue $0$,
and $\{3\}$ and $\{4\}$, corresponding to the eigenvalues $1$ and $-1$, respectively.  The last two are travelled with equal probabilities $\omega_3$ and $\omega_4$,
Thus, whatever the interpretation of the results reported in \cite{CAT}, they correspond to two completely different statistical ensembles, only one of which can be "shaped" out of the original quantum system at any given time. Hence no paradox.
 \newline
Familiar with this type of reasoning, the authors of \cite{CAT} agree that 
 if the SM of $\Pp_R$ and $\si_{R}$, are carried out at the same time the "paradox" evaporates, destroyed by the mutual disturbance affecting both measurements. To avoid the disturbance,  they introduce the left-side projector  $\Pp_L=1-\Pp_R$, and perform simultaneous weak measurements of the four operators involved. It is the suggested use of VM which is of  interest for us.
 The results of the measurements are
  \begin{eqnarray}\label{z2}
\la f\ra^{weak}_{\Pp_R}=\alpha_3+\alpha_4=0,\n
\la f\ra^{weak}_{\Pp_L}=\alpha_1+\alpha_2=1,\n
\la f\ra^{weak}_{\si_R}=\alpha_3-\alpha_4=1,\n
\la f\ra^{weak}_{\si_L}=\alpha_1-\alpha_2=0, 
\end{eqnarray} 
and it is also found that $\la \La\ra^{weak}_{\Pp_R}=\la  \La\ra^{weak}_{\Pp_L}=\la  \La\ra^{weak}_{\si_R}=\la  \La\ra^{weak}_{\si_L}=0$.
It is claimed then in \cite{CAT} that the weak values (\ref{z2}) tell us that {\it "the photon is in the left arm, while the angular momentum is in the right arm"}. But this claim is unwarranted. All that we may learn from (\ref{z2}) is that the relative amplitudes $\alpha_i$ are real, and $\alpha_1=\alpha_2=\alpha_3=-\alpha_4=1/2$. In accordance with standard quantum mechanics, this allows us to predict the results of the SM in Eq.(\ref{b2}), should they be made (for example, from (\ref{z2}) we know that $\la f\ra^{strong}_{\si^\pm_R}=\omega_{3,4}=1/4 $). It by no means serves as a proof that the results of the SM  of $\Pp_R$ and $\si_R$,  discussed above,  in  "exist", any sense, simultaneously.
 \section{Weak measurements and  counterfactual statements}
 The " quantum Cheshire cat paradox", as well as other counterfactual quantum "paradoxes"  \cite{Ah3}, \cite{AhH}, \cite{DS4} are easily dismissed if one is content to see 
 a quantum system as a  "toolbox" from which one may assemble different classical statistical ensembles by performing different measurements at different times \cite{DS2}. 
 \newline
As yet another example of this kind, consider the "3-box" case \cite{AhM}, \cite{Ah3}, where a three-state system can reach the final state via three routes defined by intermediate projections on three states $|i\ra$, $i=1,2,3$. The initial and final states are chosen so that  
 $\alpha_1=\alpha_3=1$ and $\alpha_2=-1$.
 A strong measurement of the projector $\hat{P}_1=|1\ra\la1|$ creates two real pathways, one consisting of the route $1$, and the second one comprising routes $\{2\}$ and $\{3\}$, with the special property that it is never travelled since 
 $\alpha_2+\alpha_3=0$.
 Thus, at $t=T/2$, the system is always found in the state $|1\ra$. Similarly, a strong measurement of $\hat{P}_3=|3\ra\la3|$ creates two {\it different} pathways, with the property that the intermediate state of the system  is now $|3\ra$.  Since the pathways are different, arguing that the system is { in two different states at the same time} is no more meaningful than to use  the same piece of plasticine to make first a ball, then a cube, and later argue that { a body can be round and rectangular at the same time}.
 \newline
As in the previous Section, the authors of \cite{AhH} suggest that the use of WM allows one to test the two occurrences simultaneously - {\it "to test - to some extent - assertions that have been otherwise regarded as counterfactual."} Again, the question is to which extent? Let us measure $\hat{P}_1$ and $\hat{P}_2$ weakly, and obtain the mean pointer readings of $1$ in both cases. 
We have, therefore, learnt that the relative amplitudes $\alpha_1$ and $\alpha_3$ in Eq. (\ref{a8}) both have values of 1. 
Also, since $\sum_i\alpha_i=1$, we conclude that $\alpha_2=-1$. But  we already know that, since we have prepared the transition in this particular way. As in the previous Section, we have measured the three amplitudes, and found them consistent with
what would happen in two different experiments involving strong measurements of  $\hat{P}_1$ and $\hat{P}_2$.
This is, of course, in line with the rule elementary quantum mechanics provides for assigning probabilities to exclusive alternatives \cite{FEYN}.
Again, we have found no evidence that the two strongly measured properties are possessed by the system at the same time,
just that the system can be manipulated to give the discussed results in different circumstances.
 \section{Nature's own "weak measurement"}
 The WM interference mechanism is by no means unique to the von Neumann measurements performed on 
pre- and post-selected systems, and our conclusions apply in those cases as well. To demonstrate this we consider the case of apparently "superluminal" wave packet tunnelling analysed in detail in Ref. \cite{ANN}. A particle of a mass $\mu$ described by a Gaussian wave packet with a coordinate width $\Delta x$ and mean momentum $p$,
$\psi(x,t=0)=(\Delta x)^{-1/2}G_0(x/\Delta x)\exp(ipx)$, $G_0(x)=(2/\pi)^{1/4}\exp(-x^2)$ is incident on a potential barrier placed sufficiently far to the right, 
whose transmission amplitude is $T(p)$. The energy of the particle is, $E(p)=p^2/\mu$, its free velocity is $v=p/\mu$, and we will ignore the spreading of the wave packet during the time of experiment (the case of spreading is analysed, e.g., in Ref. \cite{ANN}). As $t\to \infty$, the transmitted part of the wave function is $\psi^T(x,t)=\int T(k)G(k-p)\exp[ik(x-ct)]dk$, where $G(x)=(2pi)^{-1/2}\int G(k)\exp(ikx)dk$. Rewriting the Fourier transform as a convolution in the coordinate space we obtain an equation, similar to (\ref{a4})
 \begin{eqnarray}\label{y1}
\psi^T(x,t)=\exp(ipx-vpt/2)\int G(x-vt-x')A(x')dx',\q\q 
\end{eqnarray}
where
  \begin{eqnarray}\label{y2}
A(x)=(2\pi)^{-1/2}\exp(-ipx)\int T(k)\exp(ikx)dk.\q
\end{eqnarray}
Apart from an overall phase factor, Eq.(\ref{y1}) 
has the same form as (\ref{a4a}), with the discreet variable $S_i$ is replaced by a continuous $x'$. 
The transmitted wave packet is now built from the envelopes $G(x-vt -x')$, shifted
forwards relative to the freely propagating envelope $G(x-vt)$ if $x'>0$, or backwards of $x'<0$. The amplitude $A(x')$ for each shift 
is determined by the barrier and the incident mean momentum. We have, therefore, a "quantum measurement"  of the spacial shift $x'$ with which the transmitted particle emerges from the barrier, performed to the accuracy determined by the wave packet's width $\Delta x$.
Notably, since the particle's own position plays the role of the pointer, the "measurement" requires no external meter, and "is made" every time a wave packet is sent towards the barrier.
\newline
Our aim is to compare the mean position $\la x\ra$ of a tunnelled particle with that of freely propagating one at some large $t$,
and learn something about how long it takes for a particle to tunnel.
To ensure that the particle does not go over the barrier, we must choose $E(p)$ below the barrier height, and $\Delta x$ very large, thus making the "measurement" weak. Proceeding as in Sect.V we easily find \cite{ANN}
  \begin{eqnarray}\label{y3}
\delta x\equiv \la x\ra-vt=\int dx' x' \R \alpha(x'), \n \alpha(x)\equiv A(x)/\int dx' A(x')dx'
\end{eqnarray}
while the change in the mean momentum $\la k \ra$ of the transmitted particle, obtained because higher momenta tunnel more easily, 
is given by \cite{ANN}
  \begin{eqnarray}\label{y4}
\delta k \equiv \la k \ra-p=2\int k^2 |G(k)|^2 dk \int dx' x' \I \alpha(x').\q
\end{eqnarray}
Both quantities are readily expressed in terms of $T(p)=|T(p)|\exp[i\Phi(p)]$, 
  \begin{eqnarray}\label{y5}
\delta x =\partial_p \Phi(p),\q \delta k =2\int k^2 |G(k)|^2 dk\n 
\times \partial_p \log |T(p)|.
\end{eqnarray}
For a broad rectangular barrier of a width $d$ and a hight $V$ we have \cite{ANN} $T(p)\sim \exp[(q-ip)d]$ with 
$q\equiv \sqrt{2\mu(V-E(p)}>0$, resulting in $\delta x \sim -d <0$.
At this point we note that since a rectangular barrier does not support bound states, $T(k)$ cannot have poles in the upper half of the complex $k$-plane, and $a(x<0)\equiv 0$ \cite{ANN}. Thus, $\delta x$ is one of those weak values which lie outside the domain where the measured quantity, in our case the shift, normally has its values.
\newline
So what have we learnt about the delay a particle experiences during a tunnelling transmission?
Firstly, as always is the case with the WM, the simple mathematics is correct, and, secondly, the effect can be observed (see, for example \cite{NIM}). Interpretation of the result (\ref{y3}) is a different matter. Having found, on average,  the tunnelled particle ahead of the freely propagating one, one is tempted to conclude that the former has spent in the barrier a time, shorter by $\delta \tau = \delta x/v$.
If so, the time its has spent in the barrier is $\tau=d/v-\delta \tau$, known also as the "phase time" \cite{REV}. But $\tau$ is almost zero for a broad rectangular barrier, and this, perhaps, means that the particle's speed under the barrier exceeds the speed of light $c$. A long standing discussion of whether a particle in a classically forbidden region may cheat Einstein's relativity can be found, for example, in Refs. \cite{REV}.
\newline
We recognise all this as yet another WM "paradox" entirely of one's own making. The notions of "time spent in the barrier" and "under-barrier velocity", as  used above,  are vague, and have no clear definition in quantum theory. In fact, we have decomposed the transmission amplitude  which determines the transition of the particle from one side of the barrier to the other into a set of sub-amplitudes
[note that $\int dx A(x) = (2\pi)^{1/2}T(p)$]. We then learned something about the real and imaginary parts of their weighted sum 
$\int dx' x' \alpha(x')$, and that is all. 

 \section{Summary and discussion}
 In summary, the phenomenon of weak measurements has a simple explanation within elementary quantum  mechanics. A quantum system at a given time is fully characterised by a vector in its Hilbert space, $|\psi\ra$ or, more precisely, by the complex amplitudes $\la i|\psi\ra$, yielding its wave function in a representation corresponding to the basis $\{|i\ra\}$.
 \newline
 In a similar way, a system making a transition between the states $|\psi\ra$ and $|\phi\ra$ is characterised by virtual (Feynman) paths connecting the two states, and the probability amplitudes ascribed to them. In case of a single intermediate measurement, 
 these amplitudes are given by $\la \phi|i\ra\la i|\psi\ra$, in the eigen basis $\{|i\ra\}$ of the measured quantity. An inaccurate weak measurement does not destroy interference between the paths and, in the absence of probabilities, determines the values of the relative amplitudes, 
 $\alpha_i=\la \phi|i\ra\la i|\psi\ra/\sum_{i'}'\la \phi|i'\ra\la i'|\psi\ra$ or, more generally, of their linear combinations.
\newline
One can be forgiven for thinking that recently the WM  have been granted more importance than they probably deserve \cite{BBC}, \cite{DMAIL}.
The possible over-interpretation of the WM results began with the original paper \cite{Ah}, whose authors were vague on the nature of  the weak values 
, qualifying them simply as a "new kind of quantum variable". They also lead the reader to believe that their result was "unusual", while the measured quantity is just the difference between the amplitudes $\alpha_i$,  which naturally become large in the case of an improbable transition, $\sum_{i'}'\la \phi|i'\ra\la i'|\psi\ra\to 0$. 
\newline
The same vagueness is responsible for the notion that the WM can help resolve counterfactual "paradoxes", since different weak values can be observed simultaneously \cite{CAT}, \cite{AhH}. In reality the WM only allow one to reconstruct the amplitudes $\alpha_i$, which can later be used to predict which happens to a system under different  incompatible conditions, and by no means suggest that  these conditions are, in some sense, realised at the same time.
\newline
Another example of  over-interpretation of WM results is the identification of the "phase time" with the mean time a tunnelling particle spends in the barrier. This leads to a conflict with Einstein's causality, which is, however, easily resolved by realising that all that is measured is only a particular integral involving the sub-amplitudes into which the transmission amplitude is partitioned. 
\newline
Finally, while certain amount of publicity given to a subject can be beneficial (notably, to the present author, who would otherwise have to find a different topic), one would like to avoid references to "surreal elements of quantum mechanics" \cite{Stein}, "electrons with disembodied charge" \cite{CAT}, a "weird quantum phenomenon known as the 'Cheshire Cat' effect" \cite{DMAIL}, or "violation of relativistic causality" \cite{NIM}, unless absolutely necessary. Conventional 
quantum mechanics provides a way around these exotic suggestions, by offering a simple, albeit by far less intriguing explanation of the WM phenomena. 

Our discussion may appear unbalanced if we do not consider also what good can, and has been, achieved with the help of the WM scheme. 
It is well known that quantum perturbation theory results contain information about the phase of the relevant amplitudes and wave functions.  Consider, for example a spin-1/2 in a state $|\psi\ra =a|1\ra+b|2\ra$. Projection on the state $|1\ra$ succeeds with the probability $P=|a|^2$, and repeating it we can determine both $|a|$ and $|b|=\sqrt{1- |a|^2}$. We may then add a small magnetic field along the $x$-axis, so that $|\psi\ra$ evolves into $(1-i\gamma \sigma_x)|\psi\ra$. The probability for a successful projection is now 
$P+\delta P$, where $\lim_{\gamma \to 0}\delta P/\gamma$ is easily found to be $2|a||b|\sin \Delta \varphi$, and  $\Delta \varphi$ is the phase difference between 
$a$ and $b$. Thus, response of the system to a small perturbation allows one determine, in an indirect way, its wave function.

The WM scheme offers a similar possibility. Let us post-select the spin in a known state $|\phi\ra$, so that the two states are connected by two virtual paths with the relative amplitudes $\alpha_i=\la \phi|i\ra\la i|\psi\ra/\la\phi|\psi\ra$, ($\hat{H}=0$). Employing a weakly coupled meter, one can determine the real and imaginary parts of $\alpha_i$, as discussed in Sects. V an IV. Then, since $\la \phi|i\ra$ are known, one determines the values of $a=\la 1|\psi\ra$ and $b=\la 2|\psi\ra$ up to an unknown constant $\la\phi|\psi\ra$, and with them the all of the state $|\psi\ra$. 

This was the strategy followed by the authors of \cite{Nature} in the more interesting case of infinite number of dimensions 
, who showed that also in this case the wave function can, in principle, be extracted from the amplitudes $\alpha_i$.
[Note, however, a mistake in Eq.(6) of \cite{Nature}: the first factor in the numerator depends on $a$, and should appear inside the sum in Eq.(7), making the suggested connection between the wave function and the WV of the projector much less direct. Rather, the said WV comes out proportional to the corresponding relative amplitude $\alpha$, which is one of the main points we make in this paper.]

 It should not then come as a surprise that quantities such as the gradient of its phase, representing local velocity in Bohm's version of quantum mechanics
\cite{Bohm},\cite{Holl}, can also be "measured" indirectly in a WM scheme. A measurement of its photonic equivalent, 
the Pointing vector was reported in \cite{Scie}, whose results were further interpreted in \cite{NJP}. Both experiments represent considerable technological advances, but do not, we argue, benefit from the exotic notions mentioned at the beginning of this Section.
In their context a "weak measurement" should mean only "a perturbative scheme where the observed mean value of an additional degree of freedom can be expressed in terms of the amplitudes on the virtual paths connecting the initial and final states of the studied system".  Which is also the precise meaning of the title given to this paper.
 \section {Acknowledgements}  Support of the Project Grupos Consolidados
UPV/EHU del Gobierno Vasco (IT-472-10) and the MINECO
Grant No. FIS2012-36673-C03-01, as well as useful discussions with Prof. E. Akhmatskaya are gratefully acknowledged.
 
\end{document}